\documentclass[fleqn,twoside]{article}
\usepackage{gc}
\usepackage{latexsym,amsmath}
  \relax

\catcode`@=11 \catcode`!=11

\expandafter\ifx\csname fiverm\endcsname\relax
  \let\fiverm\fivrm
\fi
  
\let\!latexendpicture=\endpicture 
\let\!latexframe=\frame
\let\!latexlinethickness=\linethickness
\let\!latexmultiput=\multiput
\let\!latexput=\put
 
\def\@picture(#1,#2)(#3,#4){%
  \@picht #2\unitlength
  \setbox\@picbox\hbox to #1\unitlength\bgroup 
  \let\endpicture=\!latexendpicture
  \let\frame=\!latexframe
  \let\linethickness=\!latexlinethickness
  \let\multiput=\!latexmultiput
  \let\put=\!latexput
  \hskip -#3\unitlength \lower #4\unitlength \hbox\bgroup}

\catcode`@=12 \catcode`!=12

\input{pictex.tex}

\catcode`@=11 \catcode`!=11
  
\let\!pictexendpicture=\endpicture 
\let\!pictexframe=\frame
\let\!pictexlinethickness=\linethickness
\let\!pictexmultiput=\multiput
\let\!pictexput=\put

\def\beginpicture{%
  \setbox\!picbox=\hbox\bgroup%
  \let\endpicture=\!pictexendpicture
  \let\frame=\!pictexframe
  \let\linethickness=\!pictexlinethickness
  \let\multiput=\!pictexmultiput
  \let\put=\!pictexput
  \let\input=\@@input   
  \!xleft=\maxdimen  
  \!xright=-\maxdimen
  \!ybot=\maxdimen
  \!ytop=-\maxdimen}

\let\frame=\!latexframe

\let\pictexframe=\!pictexframe

\let\linethickness=\!latexlinethickness
\let\pictexlinethickness=\!pictexlinethickness

\let\\=\@normalcr
\catcode`@=12 \catcode`!=12

\renewcommand{\theequation}{\thesection\arabic{equation}}
\newcommand{\be}{\begin{equation}}
\newcommand{\ee}{\end{equation}}
\newcommand{\beqs}{\begin{eqnarray}}
\newcommand{\eeqs}{\end{eqnarray}}
\newcommand{\la}{\label}
\newcommand{\ii}{\'{\i}}
\newcommand{\La}{\Lambda}
\newcommand{\lam}{\lambda}
\newcommand{\m}{\mu}
\newcommand{\n}{\nu}
\newcommand{\vf}{\varphi}
\newcommand{\pdr}{\partial}
\newcommand{\lp}{\left(}
\newcommand{\rp}{\right)}
\newcommand{\lb}{\left[}
\newcommand{\rb}{\right]}
\newcommand{\rmd}{\mbox{d}}
\newcommand{\rme}{\mbox{e}}

\heads{J. Quiroga {\it et al}}
      {Role of the Dilatonic potential to the Quantum
stabilization of dilatonic Anti-de Sitter Universe}

\begin{document}
\twocolumn[ \Arthead{0}{0}{0}{0}{0}

\Title{Role of the Dilatonic potential to the Quantum
stabilization of dilatonic Anti-de Sitter Universe}

\Author{J. Quiroga Hurtado\foom 1}{Department of Physics, Universidad Tecnologica de
Pereira, Colombia}{and Lab. for Fundamental Study, Tomsk State
Pedagogical University, Tomsk
634041, Russia}\\

\Author{H. I. Arcos}{Department of Physics, Universidad Tecnologica de
Pereira, Colombia}{and Instituto de F\ii sica Te\'orica, Universidade Estadual Paulista, Sao Paulo, Brazil}\\

\Author{William Ardila Urue\~na}{Department of Physics, Universidad Tecnologica de
Pereira, Colombia}

\Abstract{Quantum effects lead to the annihilation of AdS Universe
when dilaton is absent. We consider here the role of the form for
the dilatonic potential in the quantum creation of a dilatonic AdS
Universe and its stabilization. Using the conformal anomaly for
dilaton coupled scalar, the anomaly induced action and the
equations of motion are obtained. Using numerical methods the
solutions of the full theory which correspond to quantum-corrected
AdS Universe are given for a particular case of dilatonic
potential.}

          ]
\email 1{jquiroga@tspu.edu.ru}

\section{INTRODUCTION}

Recent research on Anti-de Sitter spacetime has been motivated by
AdS/CFT duality conjecture (for a review see \cite{review}).
According with this conjecture, it is possible to obtain a
correspondence between classical properties of AdS spacetime and
the properties of some conformal field theory in low dimensions.

From a different point of view, quantum field theory on curved
spaces (see \cite{BOS} for further information) provides some
hints towards our understanding of the origin of the early Universe.
One of the most successful inflationary models (anomaly induced
inflation \cite{Starobinskyetc}) is based on quantum creation of a
de Sitter universe through quantum matter effects. It is
interesting, then, to understand if this quantum creation is a
sole property of de Sitter spacetime or if it is a general result
which occurs in spaces with constant curvature.\\
In this work, we consider the role of quantum matter, interacting
with the dilaton, on the AdS universe stabilization. It is already
known that quantum effects destroy and turn AdS universe unstable
\cite{brevik}. In this work, these quantum effects appear on the
dilatonic induced conformal  anomaly in four dimensions (see
\cite{odintsov}). It is worth mentioning, that this dilatonic
induced conformal anomaly may have an holographic origin within a
Yang-Mills $N=4$ conformal supergravity (via AdS/CFT correspondence) as is shown
in \cite{SN}
\section{Description of the model}\la{descr}

Our test model is based on a four dimensional AdS space-time ($AdS_4$) with metric given by
\be
ds^2=e^{-2\lambda
\tilde{x_3}}(dt^2-(dx^1)^2-(dx^2)^2)-(d\tilde{x}^3)^2\,,
\label{metr}
\ee
with an effective cosmological constant $\La = -\lambda^2$.

This metric can be rewritten into a plane-conformal one by using the following coordinate transformation
\be
y= x^3=\frac{e^{\lambda \tilde{x_3}}}{\lambda}. \la{eq2}
\ee
Replacing $a=e^{-\lambda \tilde{x_3}}=1/(\lam x^3)=1/(\lam y)$, the metric takes the final form
\be
ds^2=a^2(dt^2-dx^2)= a^2\eta_{\m\n}dx^\mu dx^\n .
\la{metric}
\ee

In order to obtain the effective action for a scalar field coupled with
the dilaton we take the following anomaly (for a review see
\cite{Duff94,odintsov} ) \beqs
T&=&b\left( F+\frac{2}{3}\,\Box R\right)+b'G+b''\,\Box R\,+\nonumber\\
& &+a_1\frac{[(\nabla f)(\nabla f)]^2}{f^4}+a_2 \Box
\left(\frac{(\nabla f)(\nabla f)}{f^2}\right),
\qquad\la{anom}\eeqs where $b$ and $b'$ are constants, \beqs
b&=&\frac{1}{120(4\pi)^2}\;,\qquad
b'=-\frac{1}{120(4\pi)^2}\;,\nonumber\\
a_1&=&\frac{1}{32(4\pi)^2}\;\;,\qquad a_2=\frac{1}{24(4\pi)^2}\;,
\eeqs while $F$ is the square of the Weyl tensor in four
dimensions and $G$ is the Gauss-Bonnet invariant
\cite{odintsov,brevik}.

Here, it must be clear that $b''$ is an arbitrary parameter which
may vary depending on the value of the finite gravitational
renormalization. From now on, we will take $b''=0$ since this has
no physical consequences. The terms multiplying $a_1$ and $a_2$
give the dilaton contribution, where $f$ is an arbitrary function
of a scalar quantum field $\varphi$.

It can be further noticed that if we omit the dilaton terms in (\ref{anom}) the well known conformal anomaly is obtained.

\section{Effective action and field equations}

Using the conformal anomaly (\ref{anom}), the effective action can
be easily constructed \cite{ReigertTseytlinetc}. Taking into
account that we are dealing with a quantum-induced AdS space-time,
we can use a metric similar to (\ref{metric}), as it is usually
done, that is to say $g_{\mu\nu}=\rme^{2\sigma(y)}\eta_{\mu\nu}$.
Here $\eta_{\mu\nu}$ denotes the Minkowski metric. We now apply
the techniques of ref.\ \cite{ReigertTseytlinetc} to obtain the
anomaly induced effective action. The result is
\beqs
W\!\!\!&=&\!\!\!\int\rmd^4\!x{[2b'\sigma\Box^2\sigma
-2(b+b')(\Box\sigma}+\nonumber\\
& &+\eta^{\mu\nu}
(\partial_{\mu}\sigma)(\partial_{\nu}\sigma))^2+ a_1\frac{[(\nabla f)(\nabla f)]^2}{f^4}+\nonumber\\
& &+a_2 \Box \left(\frac{(\nabla f)(\nabla f)}{f^2}\right)]\;.
\eeqs
Since $\sigma$ depends on $y$ only, and supposing that $f=\varphi$,
we can simplify this last equation in order to obtain a final form for the effective action
\beqs
\!\!\!\!\!\!\!\!\!\!\!\!\!\!\!\!\!\!\!\!\!\!\!\!
 W&=&V_3\int\rmd
y{[2b'\sigma\sigma''''
-2(b+b')(\sigma''+(\sigma')^2)^2 +}\nonumber\\
 &+& a_1\frac{(\varphi')^4}{\varphi^4}\sigma + a_2\lb\frac{(\vf')^2}{\vf^2}\rb''\!\!\!\sigma + a_2\frac{(\vf')^2}{\vf^2}(\sigma')^2]\quad\;\;\;.
\eeqs
Here, $\sigma'=\rmd\sigma/\rmd y$. Now, it must be taken into account that the {\it total} effective action is formed from $W$ plus a conformal invariant functional. In the plane conformal case, as the one being considered, this conformal invariant functional is a non-important constant. Only if we consider a periodicity of some coordinates this constant becomes important, as some kind of Casimir energy which depends on the radius of the compactified dimension.
In order to take into account the effects of quantum matter in the AdS universe, we must add to the anomaly induced action the classical gravitational action. In our case the classical action includes a dilatonic potential and a kinetic term for the dilaton. We write then
\beqs\la{scl}
S_{\rm cl}&=&-\frac{1}{\kappa}\int\rmd^4\!x\,\sqrt{-g}\,(R+\frac{\beta}{2}g^{\m\n}\pdr_\m\vf\pdr_\n\vf+V(\vf)+\nonumber\\
& & +6\Lambda)=\nonumber\\
&=&-\frac{1}{\kappa}\int\rmd^4\!x\,\rme^{4\sigma}
(6\rme^{-2\sigma}((\sigma')^2+(\sigma''))+\nonumber\\
& & + \frac{\beta}{2}\rme^{2\sigma}(\vf')^2+V(\vf)+6\Lambda)\;.
\eeqs
The sum of the classical action and the effective quantum action
fully describes the dynamics of the entire quantum system.

The field equations for the action $S_{cl}+W$, are obtained by
calculating its variation with respect to $\sigma$ and $\vf$,
assuming that $\sigma$ and $\vf$ depend on the conformal variable
$y$ only. In our case the potential will be defined as
$V(\vf)=\frac{\alpha_1}{\vf}$. Performing a similar analysis as in
\cite{quiroga}, we may obtain the equations of motion in terms of
the $z$ variable , which is defined by the transformation,
\be
\rmd z=a(y)\,\rmd y\;.
\ee
The so obtained field equations have the form
\beqs
&&-4b[a^3\stackrel{....}{a}+3a^2\stackrel{...}{a} + a^2(\ddot{a})^2-5a(\dot{a})^2\ddot{a}]+\nonumber\\
&&+24b'a\dot{a}^2\ddot{a}-\frac{24}{\kappa}a^4\La - \frac{12}{\kappa}\lp a^3\ddot{a}+a^2\dot{a}^2 \rp +\nonumber\\
&&+3\beta a^7\dot{\vf}^2 + 4\alpha_1\frac{a^4}{\vf}+a_1 a^4\frac{\dot{\vf}^4}{\vf^4}+ 2a_2a^2[a^2\frac{\ddot{\vf}^2}{\vf^2}+\nonumber\\ \label{ecmov2}
&&+3a\dot{a}\frac{\dot{\vf}\ddot{\vf}}{\vf^2}+a^2\frac{\dot{\vf}\stackrel{...}{\vf}}{\vf^2} - 5a^2\frac{\dot{\vf}^2\ddot{\vf}}{\vf^3}-3a\dot{a}\frac{\dot{\vf}^3}{\vf^3}+\nonumber\\
&&+3a^2\frac{\dot{\vf}^4}{\vf^4}]=0
\eeqs
\beqs \nonumber
 {   }& &\nonumber\\
&&6a_1\ln a a^2\lp a\frac{\dot{\vf}^4}{\vf^5}-a\frac{\dot{\vf}^2\ddot{\vf}}{\vf^4}-\frac{\dot{\vf}^3}{\vf^4}\rp -2a_1a^2\dot{a}\frac{\dot{\vf}^3}{\vf^4}+\nonumber\\
&&+a_2(a^2\ddot{a}\frac{\dot{\vf}^2}{\vf^3}-a^2\ddot{a}\frac{\ddot{\vf}}{\vf^2}+ a\dot{a}^2\frac{\dot{\vf}^2}{\vf^3}-a\dot{a}^2\frac{\ddot{\vf}}{\vf^2}+\nonumber\\
&&-a^2\stackrel{...}{a}\frac{\dot{\vf}}{\vf^2}-4a\dot{a}\ddot{a}\frac{\dot{\vf}}{\vf^2}- \dot{a}^3\frac{\dot{\vf}}{\vf^2})- \frac{\alpha_1}{2}\frac{a^3}{\vf^2}+ \nonumber\\
&&-\frac{\beta}{2}a^6(a\ddot{\vf}+\dot{a}\dot{\vf})-3\beta
a^6\dot{a}\dot{\vf}=0  \la{ecmov1}
\eeqs
Here $\dot{a}=\rmd a/\rmd z$ and $\dot{\vf}=\rmd \vf/\rmd z$. As
it is argued on ref.~\cite{brevik}, it is desirable to look for
solutions of the form
\be \la{ansats} a(z)\simeq a_0\,\rme^{Hz}\;,\qquad
\varphi(z)\simeq\varphi_0\,\rme^{-\alpha Hz}\;.
\ee
Replacing this ansatz into (\ref{ecmov2}) and (\ref{ecmov1}) it can be seen that the logarithmic
term of the second equation becomes of the order of $\ln a\sim  Hz$ and
since $H$ is of the order of Plank mass (ref.~\cite{brevik99})
this term can be ignored. Furthermore, and for simplicity, we are
going to take the simplest case when the kinetic term for the
dilaton is zero, that is $\beta=0$. Then, we obtain the following
equations for $\alpha$ and $H$ \beqs
24b'H^4 - 24\frac{\La}{\kappa} - \frac{24}{\kappa}H^2 +4\frac{\alpha_1}{\vf_o}\rme^{\alpha Hz}+a_1\alpha^{4}H^{4}&=&0\nonumber\\
2a_1\alpha^{3}H^{4}+3a_2\alpha
H^{4}-\frac{\alpha_1}{2}\frac{\rme^{\alpha Hz}}{\vf_o}&=&0\;.\nonumber\\
\la{ecmovb}\eeqs

Taking $\alpha$ and $\alpha_1$ as small parameters, we can neglect
the $\alpha\alpha_1H$ terms. Equations (\ref{ecmovb}), can be then
written as
\beqs
6b'H^4 - \frac{6}{\kappa}H^2 -
6\frac{\La}{\kappa} +
\frac{\alpha_1}{\vf_o}+a_1\alpha^4H^4&=&0 \nonumber\\
2a_1\alpha^{3}H^{4}+3a_2\alpha
H^{4}-\frac{\alpha_1}{2\vf_o}&=&0\,.\quad\qquad\la{ecalf1}
\eeqs
From the equations (\ref{ecalf1}) we see that the value for $H^2$
will depend on the value for the parameter $\alpha$ through the
parameter $\alpha_1$. A general solution for $H^2$ is not short
but feasible to calculate numerically.

From the set of equations (\ref{ecalf1}) we obtain (numerically)
for the dependence of $H^2$ in terms of $\alpha_1$ the curve of
the figure (\ref{fig1}). This curve was obtained by taking for
$\La$, $a_o$ and $\vf_o$ the values $-0.01$, $1.0$ and $1.0$
respectively, and $\kappa=16\pi G$, where $G=6.63\times 10^{-8}$
is the gravitational constant.
\begin{figure}[h]
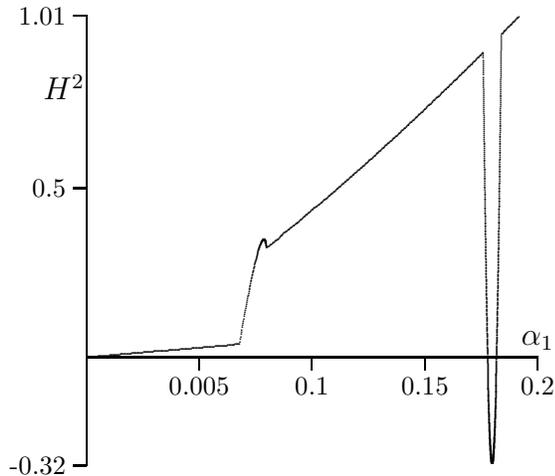

$$ \beginpicture
\setcoordinatesystem units <30cm, 45cm> \setplotarea x from 0 to
0.2, y from -0.032 to 0.101 \axis bottom shiftedto y=0 ticks withvalues
{0.005} {0.1} {0.15} {0.2}  / at 0.05 0.1 0.15 0.2 / /
\axis left shiftedto x=0 ticks withvalues {-0.32} {0.5} {1.01} / at -0.032
0.05 0.101 / /  \put {\large $\alpha_1$} at 0.2 0.005 \put {\large $H^2$} at -0.01 0.08
\setquadratic \plot  0.004 0.0002236135828
 0.008 0.0004472276102
 0.012 0.0006708422954
 0.016 0.0008944565972
 0.020 0.0011180708843
 0.024 0.0013416860609
 0.028 0.0015652994501
 0.032 0.0017889136632
 0.036 0.0020125278968
 0.040 0.0022361420952
 0.044 0.0024597562893
 0.048 0.0026833704242
 0.052 0.0029069845227
 0.056 0.0031305985734
 0.060 0.003354212616
 0.064 0.0035778265959
 0.068 0.0040250544636
 0.076 0.0306673548374
 0.080 0.0325013965356
 0.084 0.0343663140033
 0.088 0.0369301025868
 0.092 0.0388860715593
 0.096 0.0410824866719
 0.100 0.0434733639216
 0.104 0.0454861266512
 0.108 0.0478237790818
 0.112 0.0501781360015
 0.116 0.0525280357479
 0.120 0.0548750240677
 0.124 0.0571943184452
 0.128 0.059613680848
 0.132 0.0621163383638
 0.136 0.0644691585043
 0.140 0.0669338646215
 0.144 0.0694922105239
 0.148 0.0719404703788
 0.152 0.0745459257939
 0.156 0.0770493725312
 0.160 0.0796009157704
 0.164 0.0821911535795
 0.168 0.0847907654722
 0.172 0.0874371239909
 0.176 0.0900735508919
 0.180 -0.0312598638947
 0.184 0.0953961163053
 0.188 0.0981450485481
 0.192 0.1008477184695
 /
\endpicture$$
\la{fig1}
\caption{Behavior of $H^2$ in terms of $\alpha_1$. The data in the vertical axis must be corrected by a factor of $10^{-6}$ and then added to $10^{-2}$  (for example the point placed at $0.5$ is in fact the point $0.0100005$).}
\end{figure}
From this figure, we see that $H^2$ increases as the parameter
$\alpha_1$ does, and this means that the curvature rises too. As a
consequence of this the AdS universe remains to be stable and the
possibility of quantum creation of AdS universe occurs. Although
we do not plot here for negative values of $\alpha_1$, it can be
shown that for $\alpha_1<0$ the curve turns out to be a
monotonically decreasing function. This reinforces the fact that
the stability of AdS universe grows for positive values of
$\alpha_1$ only. For negative values the stability diminish as
$\alpha_1$ raises toward $0$.
It is also seen from the figure that when $\alpha_1\simeq 0.18$,
$H^2$ quickly decays to a smaller value. This may be interpreted
as a unstable point where AdS spacetime may decay to Minkowski
space. This unstable point can also be found  on the negative side
of $\alpha_1$ at $-0.18$. These unstable points could be the
result of a failure of our approximation, since $\alpha_1$ must be
considered smaller than $0.1$ for $\alpha\alpha_1H$ to be
neglected in (\ref{ecmovb}).

\section{Summary}

In \cite{quiroga} it was shown how the form of the dilatonic
potential affects the quantum creation of AdS Universe.  In the
present work it is shown the possibility of quantum creation of a
dilatonic AdS universe when the dilatonic potential is defined as a function of the form $\frac{\alpha_1}{\vf}$ and trough a numerical choice of the
values for the parameter $\alpha_1$. This choice was performed from
the analysis of the equations of motion. The total action for the
full field theory was constructed by using an anomaly induced effective
action for a dilaton coupled to a scalar field and by adding to it the classical gravitational action.

The interest for AdS spacetime is caused mainly by AdS/CFT
correspondence and nice supersymmetric properties of this
background. There was recently a lot of interest in a brane-world
approach where our observable universe is considered as brane in a
higher dimensional space. There exists some variant of this brane-world
which is based on brane quantum effects (so-called  Brane New
World)(for its formulation see \cite{NOZ,Hawking}).

On the other hand one may wonder if there are some relations
between dS and AdS spacetimes. That is to say, it is very interesting to
understand if it would be possible to obtain an AdS universe from
the dS universe. Then, the question is how does the choice of
the dilatonic potential affects the possibility of creation of
dilatonic de Sitter universe. Furthermore, one may wonder what is the answer to this question in the case of spaces within the Brane New World scenario.
This topic will be discussed elsewhere.

\Acknow {We are  grateful to Prof.S.D. Odintsov for formulation
the problem and numerous helpful discussions. We thank Prof.P.M.
Lavrov for very useful discussions. J. Q. H., H. I. A. and W. A.
U. are professors at Unviversidad Tecnol\'ogica de Pereira. This
research was supported in part by Professorship and Fellowship
from the Universidad Tecnologica de Pereira, Colombia and by
COLCIENCIAS-Colombian agency.}

\end{document}